\documentstyle[12pt]{article}
\makeatletter
\@addtoreset{equation}{section} %%% Latex Companion, p.~16
\makeatother

\makeatletter
\@addtoreset{equation}{section} %%% Latex Companion, p.~16
\makeatother

\def\appendix#1
{
\addtocounter{section}{1}\setcounter{equation}{0}%\setcounter{section}{1}
 \renewcommand{\thesection}{\Alph{section}}
 \section*{%Appendix~
 \thesection\protect\indent \parbox[t]{16.715cm}{#1}}
 \addcontentsline{toc}{section}{Appendix \thesection\ \ \ #1}
}
\def\<#1,#2>{\left\langle#1,#2\right\rangle} %% bilinear form
\newcommand{\Dirac}{{D\mkern-11.5mu/\,}} %% Dirac operator

\def\nn{\nonumber}

\def\be{\begin{equation}}
\def\ee{\end{equation}}
\def\bea{\begin{eqnarray}}
\def\eea{\end{eqnarray}}

\newcommand{\kM}{$\kappa$-Minkowski}
\newcommand{\kkP}{$\kappa$-Poincar\'{e}}
\newcommand{\al}{\alpha}
\newcommand{\bt}{\beta}

\newcommand{\x}{{\bf x}}

\newcommand{\ts}{\left(}
\newcommand{\td}{\right)}
\newcommand{\qs}{\left[}

\newcommand{\qd}{\right]}

\newcommand{\vk}{\vec{k}}

\newcommand{\vp}{\vec{p}}

\newcommand{\D}{{\mathcal{D}}}
\newcommand{\mN}{\mathcal{N}}
\newcommand{\mM}{\mathcal{M}}

\begin{document}

%%% GAC style
%\font\ninerm = cmr9
%\def\footnoterule{\kern-3pt \hrule width \hsize \kern2.5pt}
%\pagestyle{empty}

\begin{flushright}
gr-qc/0207003 \\
\end{flushright}

 \vskip 0.5 cm
\begin{center}
{\large\bf Dirac spinors for Doubly Special Relativity\\
and $\kappa$-Minkowski noncommutative spacetime}
\end{center}
\vskip 1.5 cm
\begin{center}
{\small {\bf Alessandra AGOSTINI}$^a$, {\bf Giovanni
AMELINO-CAMELIA}$^b$
and {\bf Michele ARZANO}$^c$}\\
\end{center}
\begin{center}
$^a$Dipartimento di Scienze Fisiche, Universit\`{a} di Napoli ``Federico II'',\\
and INFN Sez.~Napoli, Via Cintia, 80125 Napoli, Italy
\\
$^b$Dipartimento di Fisica,
Universit\`{a} di Roma ``La Sapienza'' and INFN Sez.~Roma1,\\
P.le Moro 2, 00185 Roma, Italy\\
$^c$Institute of Field Physics,
 Department of Physics and Astronomy,\\
 University of North Carolina,
 Chapel Hill, NC 27599-3255, USA
\end{center}

\vspace{1cm}
\begin{center}
{\bf ABSTRACT}
\end{center}
We construct a Dirac equation that is consistent with one of the
recently-proposed schemes for a ``doubly-special relativity", a
relativity with both an observer-independent velocity scale (still
naturally identified with the speed-of-light constant) and an
observer-independent length/momentum scale (possibly given by the
Planck length/momentum). We find that the introduction of the
second observer-independent scale only induces a mild deformation
of the structure of Dirac spinors. We also show that our modified
Dirac equation naturally arises in constructing a Dirac equation
in the $\kappa$-Minkowski noncommutative spacetime. Previous, more
heuristic, studies had already argued for a possible role of
doubly-special relativity in $\kappa$-Minkowski, but remained
vague on the nature of the consistency requirements that should be
implemented in order to assure the observer-independence of the
two scales. We find that a key role is played by the choice of a
differential calculus in $\kappa$-Minkowski. A much-studied choice
of the differential calculus does lead to our doubly-special
relativity Dirac equation, but a different scenario is encountered
for another popular choice of differential calculus.
 \newpage
\baselineskip 12pt plus .5pt minus .5pt \pagenumbering{arabic}

\pagestyle{plain}

\section{Introduction}
After more than 70 years of study~\cite{stachHISTO,carloHISTO} the
``quantum-gravity problem", the problem of reconciling/unifying
gravity and quantum mechanics, is still unsolved. Even the best
developed quantum-gravity theories~\cite{strings,lqg} still lack
any observational support~\cite{qgpGAC,qgpSS,qgpDA,nickcpt} and
are still affected by serious deficiencies in addressing some of
the ``conceptual issues" that arise at the interplay between
gravity and quantum mechanics\footnote{Examples of these
conceptual issues are the ``problem of time" and the
``background-independence problem"~\cite{stachashtbook}.}. One can
conjecture that the lack of observational support might be due to
the difficulties of the relevant
phenomenology~\cite{qgpGAC,qgpSS,qgpDA,nickcpt} and that the
conceptual issues might be eventually settled, but on the other
hand it is legitimate, as long as the quantum-gravity problem
remains open, to explore possible alternative paths toward the
solution of the problem. At present it is conceivable that the
empasse in the study of the quantum-gravity problem might be due
to the inadequacy of some of the key (and apparently most natural)
common assumptions of quantum-gravity approaches. Over these past
few years there has been growing interest in alternative
quantum-gravity theories, as perhaps best illustrated by studies
which take a condensed-matter perspective on the quantum-gravity
problem~\cite{volov,laugh,gac3perspe}. Another possibility,
recently proposed by one of us~\cite{gacdsr}, is the one of a new
starting point for the search of a quantum gravity: instead of
assuming that the status of Lorentz symmetry remains unaffected by
the interplay between gravity and quantum mechanics, one can
explore the possibility that the Planck length $L_p$ ($L_p \sim
10^{-33} cm$) should be taken into account in describing the
rotation/boost transformations between inertial observers. This
would amount to a deformation of special relativity, a so-called
``doubly special
relativity"~\cite{gacdsr,dsrnext,leedsr,dsrothers} (DSR), in
which, in addition to the familiar\footnote{In presence of an
observer-independent length scale the fact that our observations,
on photons which inevitably have wavelengths that are much larger
than the Planck length, are all consistent with a
wavelength-independent speed of photons must be analyzed more
cautiously~\cite{gacdsr}: it is only possible to identify the
speed-of-light constant $c$ as the speed of long-wavelength
photons.} velocity scale $c$, also a second scale, a length scale
$\lambda$ (momentum scale $1/\lambda$), is introduced as
observer-independent feature of the laws of transformation between
inertial observers. $\lambda$ can be naturally (though not
necessarily) identified with the Planck length.

 The fact that in some doubly-special-relativity scenarios
the scale $1/\lambda$ turns out to set the maximum value of
momentum~\cite{gacdsr,dsrnext} and/or energy~\cite{dsrnext,leedsr}
attainable by fundamental particles might be a useful tool for
quantum-gravity research. In particular, it appears likely
that~\cite{gacdsr,dsrnext} the idea of a doubly special relativity
may find applications in the study of certain noncommutative
spacetimes (and we will provide here more evidence in favor of
this possibility). Moreover, while the deformation is soft enough
to be consistent with all presently-available data, some of the
predictions of doubly-special-relativity scenarios are
testable~\cite{gacdsr,frangian} with forthcoming
experiments~\cite{glast}, and therefore these theories may prove
useful also in the wider picture of quantum-gravity research, as a
training camp for the general challenge of setting up experiments
capable of reaching sensitivity to very small (Planck-length
suppressed) quantum-spacetime effects.

Some of these testable predictions, which concern spin-half
particles, have been obtained at a rather heuristic level of
analysis, since, so far, no DSR formulation of spinors had been
presented\footnote{Note however that in parallel with the present
study, the issue of a description of spinors in DSR has also been
considered, from a different perspective, in
gr-qc/0207004~\cite{AKspinors}. Moreover, after the appearance of
the first version (gr-qc/0207003v1) of this manuscript, the
problem of a DSR formulation of spinors has been considered, from
yet another alternative perspective in
hep-ph/0307205~\cite{joaospinors}. Both in gr-qc/0207004 and in
hep-ph/0307205 the possible connection with \kM\ spacetime is not
considered.}. We provide here this missing element of DSR
theories.

We focus on the specific DSR scheme used as illustrative example
in the studies~\cite{gacdsr} that proposed the DSR idea, but our
approach appears to be applicable to a wider class of DSR schemes,
including the one recently proposed by Maguejio and Smolin in
Ref.~\cite{leedsr} and the wider class of DSR schemes
considered in Refs.~\cite{dsrnext,lukienewdsr}. In fact,
in all of these DSR schemes the introduction of the second
observer-independent scale relies on a nonlinear realization of
the Lorentz group in energy-momentum space: the generators that
govern the rules of transformation of energy-momentum between
inertial observers still satisfy the Lorentz algebra, but their
action is nonlinearly modified. This does not appear to be a
necessary feature of DSR theories~\cite{gacdsr}, but it does
characterize all DSR schemes so far considered, and it plays a
central role in the structure of our proposal.

A key aspect of our analysis is the fact that we first focus
on how to formulate the relevant deformed Dirac equation in a way
that involves exclusively energy-momentum space (which is the best
understood sector of these DSR theories), and then we explore the
possibility of a corresponding deformed Dirac equation in the
spacetime sector. Confirming the indications of previous heuristic
arguments~\cite{gacdsr,dsrnext}, we find that \kM\
spacetime\cite{majrue,kpoinap,gacmaj} provides a natural host for
our DSR-deformed Dirac equation. But we also show that the
presence of DSR symmetries in \kM\ is not automatic: it requires a
specific choice of the (noncommutative) differential calculus in
\kM . Our analysis shows that a generic claim that \kM\ has DSR
symmetries is incorrect. It is only once one makes a certain
choice among the possible differential calculi in \kM\ that the
DSR symmetries emerge. It had been previously shown that a DSR-deformed
Klein-Gordon equation could be obtained with two different choices
of differential calculus, but we find that the richer structure of
the DSR-deformed Dirac equation allows to select a specific choice
of differential calculus in \kM .

In the next section we start by deriving a DSR-deformed Dirac
equation in energy-momentum space. Then in Section~3 we show that
our DSR-deformed Dirac equation can be naturally introduced in
\kM\ if a certain appropriate choice of differential calculus is
adopted. In Section~4 we show that with that choice of
differential calculus ours is the only consistent Dirac equation
that can be introduced in \kM . In Section~5 we show that by
adopting another differential calculus one is then not able to
obtain a Dirac equation that is consistent with the DSR
requirements. In Section~6 we compare our results with the ones of
other deformed Dirac equations that had been previously considered
in the literature. Section~7 contains some closing remarks.

\section{Dirac spinors for Doubly Special Relativity}
Before discussing our DSR formulation of the Dirac equation we
start with a brief review of the structure of the ordinary Dirac
equation, which will provide a useful starting point for our DSR
deformation. The approach we adopt is based on the one of
Ref.~\cite{AKtechnique}. We start by introducing operators
$\overrightarrow{A}$ and $\overrightarrow{B}$ that are related to
the generators of rotations, $\overrightarrow{J}$, and boosts,
$\overrightarrow{K}$, through
\begin{equation}
\overrightarrow{A}=\frac{1}{2}(\overrightarrow{J}+i\overrightarrow{K})
\end{equation}
\begin{equation}
\overrightarrow{B}=\frac{1}{2}(\overrightarrow{J}-i\overrightarrow{K})
 \end{equation}

The usefulness of these generators $\overrightarrow{A}$ and
$\overrightarrow{B}$ is due to the familiar relation between the
Lorentz algebra and the algebra $SU(2)\otimes SU(2)$. In fact,
from the Lorentz-algebra relations for $\overrightarrow{J}$ and
$\overrightarrow{K}$ it follows that
\begin{equation}
[A_l,A_m]=i \epsilon_{lmn} A_n ~,
\end{equation}
\begin{equation}
[B_l,B_m]=i \epsilon_{lmn} B_n ~,
\end{equation}
\begin{equation}
[A_l,B_m]=0 ~ .
\end{equation}

Spinors can be labelled with a pair of numbers $(j,j')$
characteristic of the eigenvalues of $\overrightarrow{A}^2$ and
$\overrightarrow{B}^2$. In particular, ``left-handed" and ``right
handed" spinors correspond to the cases $\overrightarrow{A}^2 = 0$
and $\overrightarrow{B}^2 = 0$ respectively. Left-handed spinors
are labelled by $(\frac{1}{2},0)$ and their transformation rules
for generic Lorentz-boost "angle" (rapidity)
$\overrightarrow{\xi}$ and rotation angle
$\overrightarrow{\theta}$ are\footnote{The notation $u(\vec{p})$
is here introduced to denote the spinor wave function in
energy-momentum space. It is natural to expect that $u(\vec{p})$
could be connected by a Fourier transform to a spinor defined on a
suitable quantum spacetime. This expectation proves to be correct,
as we show in the followings sections.}
\begin{equation}
u_L \rightarrow
\exp\left(i\frac{\overrightarrow{\sigma}}{2}{\cdot}
\overrightarrow{\theta} - \frac{\overrightarrow{\sigma}}{2}{\cdot}
\overrightarrow{\xi}\right)u_L ~,
\end{equation}
where $\overrightarrow{\sigma}$ denotes the
familiar $2 {\times} 2$ Pauli matrices.
Analogously, right-handed spinors are labelled
by $(0,\frac{1}{2})$ and transform according to
\begin{equation}
u_R\rightarrow \exp\left(i\frac{\overrightarrow{\sigma}}{2}{\cdot}
\overrightarrow{\theta}+ \frac{\overrightarrow{\sigma}}{2}{\cdot}
\overrightarrow{\xi}\right)u_R ~.
\end{equation}
It is sometimes convenient to describe a spinor with space
momentum $\overrightarrow{p}$ in terms of a pure Lorentz boost
from the rest frame:
\begin{equation}
u_R(\overrightarrow{p}) =
e^{\frac{1}{2}\overrightarrow{\sigma}{\cdot}\overrightarrow{\xi}}u_R(0)=
\left(\cosh\left(\frac{\xi}{2}\right)+\overrightarrow{\sigma}{\cdot}
\overrightarrow{n}\sinh\left(\frac{\xi}{2}\right)\right)u_R(0) ~,
\label{eqn1}
\end{equation}
and
\begin{equation}
u_L(\overrightarrow{p}) =
e^{-\frac{1}{2}\overrightarrow{\sigma}{\cdot}
\overrightarrow{\xi}}u_L(0)=
\left(\cosh\left(\frac{\xi}{2}\right)-\overrightarrow{\sigma}{\cdot}
\overrightarrow{n}\sinh\left(\frac{\xi}{2}\right)\right)u_L(0)
~,\label{eqn2}
\end{equation}
where $\overrightarrow{n}$ is the unit vector in the direction of
the boost (and therefore characterizes the direction of the space
momentum of the particle) and on the right-hand sides of
Eqs.~(\ref{eqn1}) and (\ref{eqn2}) the dependence on momentum is
also present implicitly through the special-relativistic
relations\footnote{In order to render some of our equations more
compact we adopt conventions with $c \rightarrow 1$. This should
not create any confusion since in DSR the speed-of-light constant
preserves its role as observer-independent scale (but in DSR it is
accompanied by a second observer-independent scale $\lambda$) and
the careful reader can easily reinstate $c \neq 1$ by elementary
dimensional-analysis considerations.} between the boost parameter
$\xi$ and energy $E$,
\begin{equation}
\cosh \xi=\frac{E}{m} ~, \label{rapidity}
\end{equation}
 and (the ``dispersion relation") between energy and spatial momentum
\begin{equation}
   E^2=\overrightarrow{p}^2+m^2
   \label{dispersionn}
\end{equation}
for given mass $m$ of the particle.

One must then codify the fact that left-handed and right-handed
spinors cannot be distinguished at rest. One way to do
this\footnote{Since we are here only concerned with the basics of
the DSR deformation of Dirac spinors, we take the liberty to set
aside the possible phase difference between $u_R(0)$ and
$u_L(0)$.} relies on the condition $u_R(0) = u_L(0)$, from which
it follows that
\begin{equation}
\left(\begin{array}{cc}   -I & F^{+}(\xi)\\   F^{-}(\xi) &
-I \end{array}\right) \left(\begin{array}{c}    u_R(\overrightarrow{p}) \\
 u_L(\overrightarrow{p}) \end{array}\right)=0 ~, \label{Dirac}
\end{equation}
where
\begin{equation}
  F^{\pm}(\xi)=2\left(\cosh^2\left(\frac{\xi}{2}\right)
  -\frac{1}{2}{\pm}   \overrightarrow{\sigma}{\cdot}\overrightarrow{n}
  \sinh\left(\frac{\xi}{2}\right)   \cosh\left(\frac{\xi}{2}\right)\right) ~.
 \end{equation}

Using Eqs.~(\ref{rapidity}) and (\ref{dispersionn}) it is easy to
establish the dependence on the particle's energy-momentum which
is coded in the $\xi$-dependence of Eq.~(\ref{Dirac}). This leads
to the ordinary Dirac equation formulated in energy-momentum space
%\footnote{The space-time formulation
%of the Dirac equation is then obtained straightforwardly via a
%Fourier transform: $(i\gamma^{\mu}\partial_{\mu}-m)\psi(x)=0$.}
\begin{equation}
   \left(\gamma^{\mu}p_{\mu}-m\right)u(\overrightarrow{p})=0
   ~, \label{odirac}
\end{equation}
where $\gamma^{\mu}$ are the familiar ``$\gamma$ matrices" and
\begin{equation}
u(\overrightarrow{p}) \equiv \left(\begin{array}{c}
  u_R(\overrightarrow{p}) \\    u_L(\overrightarrow{p})
  \end{array}\right) ~. \label{Diracbis}
\end{equation}

 The path we followed in reviewing the derivation of the
 ordinary special-relativistic Dirac equation provides a natural
 starting point for our announced deformation within the DSR framework.
 In fact, we relied exclusively on the algebraic properties of the
 generators of boosts and rotations (the properties of the Lorentz
 algebra, without making use of the specific representation of the
 generators of boosts and rotations as differential operators on
 energy-momentum space that is adopted in special relativity)
 and on Eqs.~(\ref{rapidity}) and (\ref{dispersionn}),
 the ordinary special-relativistic relations between energy
 and rapidity (boost parameter connecting to the rest frame)
 and between energy and momentum.
 The algebraic properties of the generators of boosts
 and rotations remain unmodified in the DSR scheme considered
 in Refs.~\cite{gacdsr} (and in the other DSR schemes considered
 in Refs.~\cite{dsrnext,leedsr,lukienewdsr}).
 In fact, the nonlinearity needed in order to introduce
 the second observer-independent scale is implemented by
 adopting a deformed representation as differential operators
 on energy-momentum space of the generators of boosts and rotations,
 but these deformed generators still satisfy the Lorentz algebra.
 Therefore in the derivation of the Dirac equation the only changes
 are introduced by the DSR deformations of the relations between energy
 and rapidity and between energy and momentum.
 In the DSR scheme considered in Refs.~\cite{gacdsr},
 on which we focus here, the relation between energy and momentum
 (the dispersion relation) is
\begin{equation}
 2\lambda^{-2}\cosh\left(\lambda E \right)
 - \overrightarrow{p}^{2}e^{\lambda E}
 = 2\lambda^{-2}\cosh\left(\lambda m \right)\,\, . \label{casimir}
\end{equation}
  The relation between rapidity and energy that holds in the DSR scheme
  considered in Refs.~\cite{gacdsr}, can be deduced from the structure
  of the corresponding DSR-deformed boost transformations,
which have been studied in Refs.~\cite{gacdsr,dsrnext}. Focusing
again on a pure Lorentz boost from the rest frame to an inertial
frame in which the particle has spatial momentum
$\overrightarrow{p} \equiv |\overrightarrow{p}|
\overrightarrow{n}$ one easily finds~\cite{dsrnext}
 \begin{equation}
 E(\xi)=m+\lambda^{-1} \ln\left(1 -\sinh\left(\lambda m \right)
 e^{- \lambda m}(1-\cosh\xi)\right)\,\, . \label{trans2}
\end{equation}
 Therefore the boost parameter $\xi$ can be expressed as a function of the
 energy using
\begin{equation}
   \cosh\xi=\frac{e^{\lambda E} - \cosh\left(\lambda m\right)}
   {\sinh\left(\lambda m\right)}\,\, .\label{xiofedsr}
 \end{equation}

In the DSR derivation of the Dirac equation the
Eqs.~(\ref{casimir}) and (\ref{xiofedsr}) must replace the
Eqs.~(\ref{rapidity}) and (\ref{dispersionn}) of the ordinary
special-relativistic case. All the steps of the derivation that
used the algebra properties of the boost generators apply also to
the DSR context (since, as emphasized above, the Lorentz-algebra
relations remain undeformed in the DSR scheme considered in
Refs.~\cite{gacdsr}, and in the other DSR schemes considered in
Refs.~\cite{dsrnext,leedsr,lukienewdsr}).

 We are therefore ready\footnote{We obtain here
 the DSR-deformed Dirac equation for the four-component
 spinor $u(\overrightarrow{p})$. We take some liberty
 in denoting with $u_R(\overrightarrow{p})$ two of the
 components of $u(\overrightarrow{p})$ and
 with $u_L(\overrightarrow{p})$ the remaining two components.
 In fact, especially if, as suggested in Refs.~\cite{gacdsr,dsrnext},
 the DSR deformation should rely on a noncommutative spacetime sector the
 action of ``space-Parity" transformations on energy-momentum space and
 on our spinors might involve some subtle issues~\cite{gacmaj}.
 The labels ``$R$" and ``$L$" on our DSR spinors are therefore
 at present only used for bookkeeping (they are reminders of
 the role that these components of the DSR Dirac spinor play
 in the $\lambda \rightarrow 0$ limit).} to write down
 the DSR-deformed Dirac equation:
\begin{equation}
\left(\begin{array}{cc}   -I & F^{+}_{\lambda}(E,m)\\
F^{-}_{\lambda}(E,m) & -I \end{array}\right)
\left(\begin{array}{c} u_R(\overrightarrow{p}) \\
u_L(\overrightarrow{p}) \end{array}\right) =0 \label{kDirac}
\end{equation}
 where
\begin{equation}
  F^{\pm}_{\lambda}(E,m)=\frac{e^{\lambda E}-\cosh\left(\lambda
  m \right)  {\pm}\overrightarrow{\sigma}{\cdot}\overrightarrow{n}\left(2e^{\lambda E}
  \left(\cosh\left(\lambda E \right) -\cosh\left(\lambda
  m \right)\right)\right)^{\frac{1}{2}}}{\sinh\left(\lambda
  m \right)}\,\, .
  \end{equation}

Introducing
 \begin{equation}
  D^{\lambda}_{0}(E,m) \equiv \frac{e^{\lambda E}-\cosh\left(\lambda
  m \right)}  {\sinh\left(\lambda m \right)}
  \end{equation}
  and
  \begin{equation}\label{diki}
  D^{\lambda}_{i}(E,m) \equiv \frac{n_{i}\left(2
  e^{\lambda E}\left(\cosh\left(\lambda E \right) -\cosh\left(\lambda
  m \right)\right)\right)^{\frac{1}{2}}}{\sinh\left(\lambda m \right)}
\end{equation}
the DSR-deformed Dirac equation can be rewritten as
\begin{equation}
\left(\gamma^{\mu}D^{\lambda}_{\mu}(E,m)-I\right)
u(\overrightarrow{p}) =0\label{kDirac1}
 \end{equation}
where again the $\gamma^{\mu}$ are the familiar ``$\gamma$
matrices".

The nature of this DSR deformation of the Dirac equation becomes
more transparent by rewriting (\ref{diki}) taking into account the
DSR dispersion relation (\ref{casimir}):
 \begin{equation}
  D^{\lambda}_{i}(\overrightarrow{p},m)
  =\frac{e^{\lambda E}}  {\lambda^{-1}\sinh\left(\lambda m \right)}
  p_{i} ~. \label{dikip}
\end{equation}
 In particular, as one should expect,
 in the limit $\lambda\rightarrow 0$ one finds
\begin{equation}
 D^{\lambda}_{i}(E,m)\rightarrow\frac{E}{m} ~, \label{limit1}
\end{equation}
 \begin{equation}
 D^{\lambda}_{i}(\overrightarrow{p},m)\rightarrow\frac{p_{i}}{m} ~,
\end{equation}
and the familiar special-relativistic Dirac equation is indeed
obtained in the $\lambda\rightarrow 0$ limit.

It is also easy to verify that the determinant of the matrix
$(\gamma^{\mu}D^{\lambda}_{\mu}(E,m)-I)$ vanishes, as necessary.
In fact,
\begin{eqnarray}
\det\left(\gamma^{\mu}D^{\lambda}_{\mu}(E,m)-I\right) =  \left(
\sinh^{2}\left(\lambda m \right)-  \left(e^{\lambda
E}-\cosh\left(\lambda m \right)\right)^{2}  +\frac{e^{2 \lambda
E}}{\lambda^{-2}}\overrightarrow{p}^{2}\right)^{2} =\nonumber\\
 = \left(\frac{e^{\lambda E}}{\lambda^{-2}}\left(   -
 2\lambda^{-2}\cosh\left(\lambda E \right)+  \overrightarrow{p}^{2}
 e^{\lambda E}  +2\lambda^{-2}\cosh\left(\lambda
 m \right)\right)\right)^{2}=0 ~, \label{det}
\end{eqnarray}
where the last equality on the right-hand side follows from the
DSR dispersion relation.

Our DSR-deformed Dirac equation of course leads to the
DSR-deformed Weyl equation in the case of massless particles. In
terms of the ``DSR helicity" of our massless spinors one finds:
 \begin{equation}
\left(\overrightarrow{\sigma}{\cdot}\hat{p}\right)
u_{R,L}(\overrightarrow{p})
  =   {\pm}   u_{R,L}(\overrightarrow{p})\,\, , \label{weyl}\end{equation}
 where $\hat{p} \equiv \overrightarrow{p}/|\overrightarrow{p}|$.
 The operator $\overrightarrow{\sigma}{\cdot}\hat{p}$ still has
 eigenvalues ${\pm}1$ as in the ordinary special-relativistic case.

In summary the DSR description of spinors appears to require only
a relatively mild deformation of the familiar special-relativistic
formulas. Our DSR-deformed Dirac equation differs from the
ordinary Dirac equation only through the dependence on
energy-momentum of the coefficients of the $\gamma^{\mu}$
matrices. The difference between the DSR coefficients,
$[D^{\lambda}_{0}(E,m),D^{\lambda}_{i}(\overrightarrow{p},m)]$,
and the ordinary ones, $[E/m,\overrightarrow{p}/m]$, is very small
($\lambda$-suppressed, Planck-length suppressed) for low-energy
particles, and in particular the difference vanishes in the
zero-momentum limit. Still it is plausible that the new effects
might be investigated experimentally in spite of their smallness,
following the strategy outlined in the recent
literature~\cite{qgpGAC,qgpSS,qgpDA,grbgac,gactp} on the
search of Planck-length suppressed effects.

\section{DSR Dirac Equation in $\kappa$-Minkowski spacetime}

\subsection{DSR, \kM\ and $\kappa$-Poincar\'{e} Hopf algebras}
In the previous section our analysis has been performed entirely
in the energy-momentum space. It turn out to be possible to
specify completely the DSR-deformed Dirac equation in
energy-momentum space using only the chosen DSR laws of
transformation of energy-momentum, without any assumption about
the nature and structure of spacetime. We now want to look for a
spacetime realization of our DSR-deformed Dirac equation. We will
verify, using a standard procedure~\cite{Dirac} for the
construction of a Dirac equation, that our DSR-deformed Dirac
equation is appropriate for the description of spin-$\frac{1}{2}$
particles in the \kM\ noncommutative spacetime (reviewed in the
next subsection).

We consider \kM\ spacetime because of various
indications~\cite{gacdsr,dsrnext} that the symmetries of
this spacetime may be compatible with the DSR requirements. These
indications are so far incomplete, especially for the analysis of
multiparticle systems in \kM , but at least in the one-particle
sector the presence of DSR-deformed Lorentz symmetry in \kM\ is
rather well established. In work that preceded the proposal of DSR
theories of Refs.~\cite{gacdsr}, it had already been argued that
the so-called $\kappa$-Poincar\'{e} Hopf algebra~\cite{kpoinap}
could describe deformed {\underline{infinitesimal}} symmetry
transformations for $\kappa$-Minkowski, but it was believed that
these algebra structures would not be compatible with a genuine
symmetry group of finite transformations (on the basis of a few
attempts~\cite{kpoinnogroup} it was inferred that the emerging
structure would be the one of a ``quasigroup"~\cite{batalin}, with
dubious applicability in physics). However, already in
Refs.~\cite{gacdsr}, it was observed that at least one
formulation of the  $\kappa$-Poincar\'{e} Hopf algebra did allow
for the emergence of a group of finite transformations of the
energy-momentum of a particle (while the same is not true for
other formulations, the so-called ``bases", of the Hopf algebra).
That result amounts to proving that the mathematics of the
$\kappa$-Poincar\'{e} Hopf algebra (and therefore possibly
$\kappa$-Minkowski) can meaningfully describe the one-particle
sector of a physical theory in a way that involves DSR-deformed
Lorentz symmetry. But it is still unclear whether there is a
formulation of the $\kappa$-Poincar\'{e} Hopf algebra that can be
used to construct a theory which genuinely enjoys deformed Lorentz
symmetry throughout, including multiparticle systems.

The recipe adopted in the $\kappa$-Poincar\'{e} literature for the
description of two-particle systems relies on the law of composition
of momenta obtained through a ``coproduct sum" $(p \dot{+} p')^\mu$
(where $(p \dot{+} p')^\mu = \delta^{\mu , 0} (p^0 + p'^0) +
\delta^{\mu , j} (p^j + e^{p^0 / \kappa} p'^j)$), and
the action of boosts on
the composed momenta which is
induced by the action on each of the momenta
entering the composition. This has been adopted in the
$\kappa$-Poincar\'{e} literature even very
recently~\cite{lukienewdsr}, not withstanding the new DSR-deformed
Lorentz symmetry perspective proposed in Ref.~\cite{gacdsr}. From
a DSR perspective this $\kappa$-Poincar\'{e} description of
two-particle systems is not acceptable: for a particle-producing
collision process $a+b \rightarrow c+d$ laws~\cite{lukienewdsr}
of the type $(p_a \dot{+} p_b)^\mu = (p_c \dot{+} p_d)^\mu$, are
inconsistent~\cite{gac3perspe}
with the laws of transformation for the momenta of
the four particles. In fact, one finds that the condition $(p_a
\dot{+} p_b)^\mu = (p_c \dot{+} p_d)^\mu$ can be imposed in a
given inertial frame but it will then be violated in other
inertial frames ({\it
i.e.} $(p_a \dot{+} p_b)^\mu - (p_c \dot{+} p_d)^\mu
= 0 \rightarrow (p_a' \dot{+} p_b')^\mu - (p_c' \dot{+} p_d')^\mu \neq 0$).

Therefore the possibility that there would be a formulation of
$\kappa$-Poincar\'{e} that is fully compatible with the DSR
requirements remains an open problem. And, correspondingly, it
remains to be established whether one can formulate a physical
theory in $\kappa$-Minkowski spacetime that is acceptable from a
DSR perspective. However, as mentioned, all the difficulties
appear to emerge only outside the one-particle sector (for
multiparticle processes), and it is therefore not surprising that,
as we will show, the (single-particle) Dirac equation in \kM\
spacetime turns out to be consistent with the DSR-deformed Dirac
equation we obtained in the previous section.

Interestingly our analysis suggests that the presence of DSR
symmetries in the one-particle sector of \kM\ is not automatic: it
requires a specific choice of the (noncommutative) differential
calculus in \kM . It is sometimes stated in very general terms
that the one-particle sector of physical theories in \kM\ should
enjoy DSR symmetries. This probably comes from the analysis of the
Klein-Gordon equation in \kM\ which indeed inevitably takes a
DSR-compatible form. But in our analysis of the Dirac equation in
\kM\ we find that there is a crucial choice between different
formulations of the differential calculus in \kM , and that the
DSR-deformed Dirac equation is only obtained upon making an
appropriate choice of differential calculus.

After a brief review of some important features of \kM\ spacetime,
given in the next subsection, in Subsection~3.3 we show that with
a given choice of differential calculus and a given {\it ansatz}
for the form of the Dirac equation one indeed obtains the
DSR-deformed Dirac equation. Then in Section~4 we show that the
chosen differential calculus inevitably leads to the DSR-deformed
Dirac equation (independently of any {\it ansatz}). And in
Section~5 we show that an alternative choice of differential
calculus does not give us a DSR-deformed Dirac equation.

\subsection{\kM\ spacetime}
\kM ~\cite{majrue,kpoinap} is a Lie-algebra noncommutative
spacetime~\cite{wess} with coordinates satisfying the commutation
relations \be [{\x}_0,{\x}_j]=i\lambda
{\x}_j~,~~~[{\x}_j,{\x}_k]=0 \label{eq:kM} \ee where $j,k=1,2,3$.
The noncommutativity parameter $\lambda$ has dimensions of a
length, in natural units $\hbar = c = 1$. (In most of the \kM\
literature one finds the equivalent parameter $\kappa$, which is
$\kappa = 1/\lambda$, but our formulas turn out to be more compact
when expressed in terms of $\lambda$.) Of course, conventional
commuting coordinates are recovered in the limit $\lambda\to 0$.

The elements of \kM\ (the ``functions of \kM\ coordinates") are
sums and products of the noncommuting coordinates $\x_{\mu}$. It
is possible to establish a  correspondence between elements of
\kM\ and analytic functions of four commuting variables $x_\mu$.
Such a correspondence is called ``Weyl map", and is not
unique~\cite{alz02,aad03} since it depends on an ordering choice.
For example the very simple commutative function $x_2 t$  can be
mapped into different functions in \kM : two possibilities are
${\x}_2 {\x}_0$ and ${\x}_0{\x}_2={\x}_2{\x}_0 + i \lambda {\x}_2$
(of course the ordering issue disappears in the
$\lambda\rightarrow 0$ limit, where ${\x}_2 {\x}_0 =
{\x}_2{\x}_0$).

Many properties of a noncommutative spacetime are very naturally
described in terms of a Weyl map~\cite{wess}. While, as mentioned,
different Weyl maps can be considered, in this paper for
definiteness we will work with the ``time-to-the-right-ordered
map". It is sufficient to specify this Weyl map $\Omega$ on the
complex exponential functions and extend it to the generic
function $\phi(x)$, whose Fourier transform is
$\tilde{\phi}(k)=\frac{1}{(2\pi)^4}\int d^4x\,\phi(x)e^{-ikx}$, by
linearity \be \Phi(\x)\equiv\Omega(\phi(x))=\int d^4k\,
\tilde{\phi}(k)\, \Omega(e^{ikx}) =\int d^4k \,\tilde{\phi}(k)\,
e^{-i\vk {\cdot} \vec{\x}}e^{ik_0{\x}_0} \label{kMelem} \ee
(Here and in the following we adopt conventions such
that $k x \equiv k_\mu x^\mu \equiv k_0 x^0 - \vk {\cdot} \vec{\x}$.)

On the basis of the analysis reported in Ref.~\cite{aad03} we
expect that, consistently with this choice of Weyl map,
translations, $P_{\mu}$, rotation, $M_j$, and boosts ${\mN}_j$
should be described as follows \bea
&&P_{\mu}\Phi({\x})=\Omega[-i\partial_{\mu}\phi(x)]\nn\\
&&M_j\Phi({\x})=\Omega[i\epsilon_{jkl}x_k\partial_l\phi(x)]\nn\\
&&{\mN}_j\Phi({\x})=\Omega(-[ix_0\partial_j
+x_j(\frac{1-e^{2i\lambda \partial_0}}{2\lambda}
-\frac{\lambda}{2}\nabla^2) +\lambda
x_l\partial_l\partial_j]\phi(x)) \label{bicrossbasis} \eea

These generators satisfy the requirements for the Majid-Ruegg
bicrossproduct basis of the \kkP\ Hopf
algebra~\cite{majrue,kpoinap}, with the following commutation
relations \bea
\qs P_{\mu},P_{\nu}\qd&=&0\nn\\
\qs M_j,M_k\qd&=&i\varepsilon_{jkl}M_l,\;\;\;\qs {\mN}_j,M_k\qd
=i\varepsilon_{jkl}{\mN}_l,\;\;\;\qs {\mN}_j,{\mN}_k\qd
=-i\varepsilon_{jkl}M_l\nn\\
\qs M_j,P_0\qd&=&0,\;\;\;[M_j,P_k]=i\epsilon_{jkl}P_l\nn\\
\qs {\mN}_j,P_0\qd&=&iP_j\nn\\
\qs {\mN}_j,P_k\qd&=&i\qs\ts \frac{1- e^{-2\lambda
P_0}}{2\lambda}+\frac{\lambda}{2}\vec{P}^2\td\delta_{jk} -\lambda
P_jP_k\qd\nn \eea and the following co-algebra relations \bea
\Delta(P_0)&=&P_0\otimes 1+1\otimes P_0\;\;\;\Delta(P_j)
=P_j\otimes 1+e^{-\lambda P_0}\otimes P_j\nn\\
\Delta(M_j)&=&M_j\otimes 1+1\otimes M_j\nn\\
\Delta({\mN}_j)&=&{\mN}_j\otimes 1+ e^{-\lambda P_0}\otimes
{\mN}_j-\lambda\epsilon_{jkl}P_k\otimes M_l ~. \eea

The ``mass-squared" Casimir operator in this Majid-Ruegg
bicrossproduct basis takes the form \be
C_{\lambda}(P)=\cosh(\lambda P_0)-\frac{\lambda}{2}e^{\lambda
P_0}\vec{P}^2 \ee and from this one easily obtains (see, {\it
e.g.}, Ref.~\cite{aad03}) that the Klein-Gordon equation should be
written as \be \ts
\Box_{\lambda}+M_{KG}^2\td\Phi({\x})=0\label{defKGeq} \ee in terms
of the differential operator $\Box_{\lambda}$, which is a
deformation of the familiar D'Alembert operator \be
\Box_{\lambda}=-\frac{2}{\lambda^2}\qs\cosh(\lambda P_0)-1\qd+
e^{\lambda P_0}\vec{P^2} \ee and of the ``Klein-Gordon mass
parameter" $M_{KG}$.

Making use of the (\ref{kMelem}) we can write the Klein-Gordon
equation in energy-momentum space \be \ts 2\lambda^{-2}\qs
\cosh(\lambda k_0)-1\qd- e^{\lambda
k_0}\vec{k^2}-M_{KG}^2\td\tilde{\phi}(k)=0 \ee from which it is
easy to derive the dispersion relation for a free scalar  particle
in \kM\
\bea
&&\cosh(\lambda E)-\frac{\lambda^2}{2} e^{\lambda
E}\vec{p}^2=1+\frac{\lambda^2}{2}M_{KG}^2\label{kmdisprel}~,
\eea
where $E$ and $\vec{p}$ denote the energy and the space-momentum
of the particle ($E=E(\vec{p})$ from (\ref{kmdisprel}) for
particles ``on shell"). Clearly the mass parameter $M_{KG}$
is not the rest energy, but $M_{KG}$ and the rest energy (rest
mass) $m$ are connected by the relation
$M_{KG}=\sqrt{2(\cosh(\lambda m)-1)}/\lambda$.

The relation (\ref{kmdisprel}) is the same dispersion relation
(\ref{classlim}) which we considered in Section~2 on the basis of
the DSR requirements.

\subsection{A DSR Dirac equation in \kM}
We are basically ready to investigate whether the
energy-momentum-space deformed Dirac equation obtained in
Section~2 can be seen as the energy-momentum-space counter-part of
a natural Dirac equation in \kM\ noncommutative spacetime. We
intend to follow closely the line of analysis originally adopted
by Dirac in the conventional case of commutative Minkowski
spacetime. This Dirac procedure consists in writing a partial
differential equation linear in the derivatives with arbitrary
coefficients: \be ( i
\gamma^{\mu}\partial_{\mu}+mI)\psi(x)=0\label{diraccomm} \ee where
$m$ is the particle mass, $\psi$ is an $n$-plet of fields, and the
$\gamma_{\mu}$  are $n {\times} n$ hermitian matrices to be
determined by imposing the dispersion relation (that works as
physical condition). It turns out that $n=4$ or greater is
required for consistency.

Also in \kM , as in the commutative case, we introduce Lorentz
spinor wave functions $\Psi(\x)$, whose components are of the
form: \be \Psi_r({\x})=\int d^4k \, \tilde{\psi}_r(k)
e^{-ik{\x}}e^{ik_0{\x_0}} \label{joFT} \ee
where $r$ is a
spin index. $\Psi(\x)$ will represent a physical state in \kM\
when it satisfies a wave equation
with a (deformed) $\Dirac_{\lambda}$
operator $[\Dirac_{\lambda}+M_DI]\Psi(\x)=0$.

In seeking a suitable form for $\Dirac_{\lambda}$ it is important
to notice that, while in the commutative case there is only one
natural differential calculus (which involves the ordinary
derivatives and gives rise to the Dirac equation
(\ref{diraccomm}), in the case of \kM\ the introduction of a
differential calculus is a more complex
problem~\cite{Sitarz,Masl}. As announced, in this section we focus
on one possible choice of differential calculus, the
``five-dimensional differential calculus" of Ref.~\cite{Sitarz}.
In this 5D differential calculus the exterior derivative operator
$d$ of a generic \kM\ element $F({\x})=\Omega(f(x))$ can be
written in terms of vector fields $\D_a(P)$ as follows:
 \bea
&&dF({\x})=dx^a{\D}_a(P)F({\x}),\;\;\;\;a=0,\dots,4 \label{5Ddiffcalc1}\\
&&{\D}_0(P)=\frac{i}{\lambda}[\sinh(\lambda P_0)
+\frac{\lambda^2}{2}e^{\lambda P_0}P^2] ~, \nn\\
&&{\D}_j(P)= iP_j e^{\lambda P_0} ~,
j=1,2,3 ~, \nn\\
&&{\D}_4(P)=-\frac{1}{\lambda}(1-\cosh(\lambda P_0)
+\frac{\lambda^2}{2}P^2 e^{\lambda P_0}) ~.\nn
\eea
Of course, these \kM\ derivative vector fields $\D_a$ reproduce
their commutative-Minkowski counterparts in the $\lambda \rightarrow 0$
limit:
$$
\lim_{\lambda  \rightarrow
0}\D_\mu(P)=iP_\mu=\partial_\mu,\;\;\;\lim_{\lambda  \rightarrow 0}\D_4(P)=0.
$$
The introduction of this 5D calculus in our 4D spacetime may at
first appear to be surprising, but it can be naturally introduced
on the basis of the fact that the \kkP /\kM\ framework can be
obtained (and was originally obtained~\cite{LNR92} by contraction
of a quantum-deformed anti-de Sitter algebra). The fifth one-form
generator is here denoted by ``$dx_4$", but this is of course only
a formal notation since there is no fifth \kM\ coordinate
${\x}_4$. And the peculiar role of $dx_4$ in this differential
calculus is also codified in the fact that, if one examines the
deformed derivatives (\ref{5Ddiffcalc1}) on-shell, {\it i.e.} with
$E=E(p)$ satisfying the dispersion relation (\ref{kmdisprel}) \be
\cosh(\lambda E)-\frac{\lambda^2}{2}e^{\lambda
E}\vec{p}^2=\cosh(\lambda m) \label{disprel}\ee the last component
${\D}_4(P)$ can be written as a pure function of the mass: \be
{\D}_a(E,p)=\ts\frac{i}{\lambda}[e^{\lambda E} -\cosh{\lambda
m}],ip_je^{\lambda E}, \lambda^{-1}(\cosh(\lambda m)-1)\td
\label{Donshell}\ee

The deformed Klein-Gordon
equation  (\ref{defKGeq}) takes a very simple form
in terms of
the five-dimensional differential calculus:
\be [\D^a\D_a+M_{KG}^2]\Phi(\x)=0 \label{DaDa}\ee
where $\D^a\D_a\equiv \D_0^2-\sum_{j=1}^3\D_j^2+\D_4^2$.
In fact, $\D^a\D_a=\Box_\lambda$, and therefore Eq.~(\ref{DaDa})
is equivalent to the Klein-Gordon equation (\ref{defKGeq}).

Let us now consider a Dirac equation in \kM\ spacetime of the
general form \be \ts\Dirac_{\lambda}+M_{D}I\td\Psi({\x})=0
\label{eq:kmdirac}\ee where $I$ is the identity matrix, $M_{D}$ is
a mass parameter analogous to $M_{KG}$ (like $M_{KG}$ it will be
related to the rest energy $m$) and $\Dirac_{\lambda}$ is the
deformed Dirac operator

The Dirac operator must satisfy three key requirements:\\

i) \emph{Commutative limit}: in the limit $\lambda \to 0$ one must
find that $\Dirac_{\lambda}$ reduces to the classical operator
$\Dirac=i\partial_{\mu}\gamma^{\mu}$ in terms of usual Dirac
$\gamma^{\mu}$ (${\mu}=0,...,3$) matrices.\\

ii)\emph{Physical condition}: $\Dirac_{\lambda}$ must be such that
the components of $\Psi$ must satisfy the $\kappa$-deformed KG
equation (\ref{eq:KG}), \emph{i.e.} ``plane waves on shell", with
momenta ($E,\vec{p}$) satisfying the dispersion relation (\ref{kmdisprel}),
must be  solutions of (\ref{eq:kmdirac}). The most general form of
the ``plane wave on shell" is:
\be
u(\vec{p})e^{-ip_j\x_j}e^{iE\x_0}+v(\vec{p})e^{-iS(p_j)}
e^{iS(E)\x_0}\label{decomp}
\ee
where $S(E,\vec{p})=(E,-e^{\lambda E}\vec{p})$ is the ``antipode map",
which generalizes the inversion operation in \kM . In fact
both $e^{-ip_j\x_j}e^{iE\x_0}$ and $e^{-iS(p_j)}e^{iS(E)\x_0}$ are
solutions of the $\kappa$-deformed KG equation (if $E=E(p)$
satisfies the dispersion relation (\ref{kmdisprel})). Thus, the
following equations must be satisfied:
\bea
&&(\Dirac_{\lambda}-M_DI)u_r(\vec{p})e^{-ip_j\x_j}e^{iE(p)\x_0}=0\nn\\
&&(\Dirac_{\lambda}-M_DI)v_r(\vec{p})e^{-iS(p_j)\x_j}e^{-iE(p)\x_0}=0.\nn\
\eea
We focus our attention on the equation for $u(\vec{p})$.
Then the equation for $v(\vec{p})$ will be
straightforwardly found
substituting $(E,\vec{p})$ with $(S(E),S(\vec{p}))$.  \\

iii) \emph{Covariance property}: $\Dirac_{\lambda}$ must
be invariant, $[T,\Dirac_{\lambda}]=0$, under the action of
all generators $T$ in the symmetry algebra.\\

We start by noticing that from \be \qs
\Dirac_{\lambda}(E(p),p)+M_DI\qd u(\vec{p})=0 ~, \ee one obtains,
acting with $\Dirac_{\lambda}-M_DI$, \be
[\Dirac_{\lambda}-M_DI][\Dirac_{\lambda}
+M_DI]u(\vec{p})=0\;\Rightarrow\; (\Dirac_{\lambda}^2
-M_D^2)_{on-shell}=0 \ee

Observing that the components ${\mathcal{D}}_a$ introduced
above transform under the \kkP\ action exactly as the standard
derivatives transform under the standard Poincar\'{e} action, and in
particular
\bea &&\qs N_j,{\D}_0\qd=i{\D}_j\;\; \qs
N_j,{\D}_k\qd =i{\D}_0\;\;\qs N_j,{\D}_4\qd=0 ~,
\label{covar} \eea
it appears natural to make the following {\it ansatz} for the Dirac
operator $\Dirac_{\lambda}$:
\be
\Dirac_{\lambda}(P)=i\gamma^{\mu}{\D}_{\mu}(P) \label{ansatz}
\ee
where $\gamma^{\mu}$ ($\mu=0,\dots,3$) denotes again the usual
(undeformed) Dirac  matrices. Essentially in (\ref{ansatz}) on
obtains the deformed Dirac operator using only the first four
(more familiar) components of the vectorial field $\D_a$.

Our {\it ansatz} for the Dirac operator turns out to lead to a
satisfactory Dirac equation; in fact, the requirement i) is
self-evidently satisfied, the requirement ii) is
satisfied with the condition that
 the parameter $M_D$ and the rest energy $m$
 are related by $M_{D}=\sinh(\lambda m) /\lambda$; in fact:
\bea \left[ \Dirac^2
\right]_{on shell}
&=&[i\gamma^\mu\D_\mu(E,p)]^2=-\gamma^\mu\gamma^\nu\D_\mu\D_\nu=
-\frac{1}{2}\gamma^\mu\gamma^\nu(\D_\mu\D_\nu+\D_\mu\D_\nu)\nn\\
&=&-\frac{1}{2}\{\gamma^\mu,\gamma^\nu\}\D_\mu\D_\nu=-\D_\mu\D^\mu
=\sum_j\D_j^2(E,p)
-\D_0^2(E,p)\nn \eea
and using (\ref{disprel}),(\ref{Donshell}) one finds that
 \bea
\left[ \Dirac^2 \right]_{on shell} &=&-e^{2\lambda
E}\vec{p}^2+\lambda^{-2}[e^{2\lambda E} -2e^{\lambda
E}\cosh(\lambda m)+\cosh(\lambda m)^2]\nn\\
&=&2\lambda^{-2}e^{\lambda E}[\cosh(\lambda m)-\cosh(\lambda
E)]+\lambda^{-2}[e^{2\lambda E} -2e^{\lambda E}\cosh(\lambda
m)+\cosh(\lambda m)^2]\nn\\
 &=&\lambda^{-2}\sinh(\lambda m)^2=M_D^2. \nn
\eea

Finally, the invariance requirement iii) is clearly
satisfied  in light of the covariant transformation properties
(\ref{covar}) of the $\D_a$.

We are therefore led to the following spacetime (\kM\ spacetime)
formulation of the DSR-deformed Dirac equation
\be \qs
i{\D}_{\mu}(P)\gamma^{\mu} +\frac{\sinh(\lambda
m)}{\lambda}I\qd\Psi({\x})=0 \ee
(we remind the reader that the $P_\mu$ are defined as
the differential operators of (\ref{bicrossbasis})).

The corresponding energy-momentum-space formulation
of this DSR-deformed Dirac equation can be obtained
through the Fourier transform (\ref{joFT}),
and takes the form
\be \qs i{\D}_{\mu}(k)\gamma^{\mu}+\frac{\sinh(\lambda
m)}{\lambda}I\qd\tilde{\psi}(k)=0 \label{energymomenta}\ee
Using the decomposition (\ref{decomp}) of the Dirac spinor
on shell, the equation (\ref{energymomenta}) reproduces
exactly the DSR-deformed equation (\ref{kDirac1}):
\be \qs \frac{1}{\lambda}[e^{\lambda E}
-\cosh{\lambda m}]\gamma^0+e^{\lambda E}p_j\gamma^j
-\frac{\sinh(\lambda m)}{\lambda}I\qd u(\vec{p})=0 \ee

It is already noteworthy that our energy-momentum-space
DSR-deformed Dirac equation (\ref{kDirac1}), first derived in
Section~2 from general DSR symmetry principles (without advocating
in any way properties of the \kM\ spacetime) emerges in \kM\
spacetime, upon a suitable choice of differential calculus and
within a natural {\it ansatz} for the formulation of the Dirac
equation in terms of the elements of the chosen differential
calculus. We are however hoping to establish an even more robust
connection with \kM\ spacetime, and therefore in the following
sections we examine whether our result depends crucially on the
{\it ansatz} (\ref{ansatz}) and/or on the choice of differential
calculus.

\section{Uniqueness of the Dirac equation
for given choice of differential calculus}

\subsection{Constructing the deformed Dirac equation}
The next step of our analysis relies once again on the
differential calculus (\ref{5Ddiffcalc1}), but explores the
structure of the Dirac equation in \kM\ in otherwise completely
general terms, without resorting to the {\it ansatz}
(\ref{ansatz}). If the differential calculus is given by
(\ref{5Ddiffcalc1}) the most general parametrization of the Dirac
equation in \kM\ is: \be \ts i\D_0+i\al^j\D_j+\al^4\D_4+\bt M_D\td
\Psi(\x)=0 \label{pde:dirac} \ee where $\Psi(\x)$ is again an
$n$-plet (vector) of wave functions of noncommutative coordinates
$\x$, and $\al^i$ ($i=1,2,3$), $\al^4$, $\bt$ are five hermitian
matrices. $M_D$ is once again a mass parameter which we expect to
be some simple function of the physical mass $m$.

The {\it ansatz} considered in the previous section corresponds
to $\alpha^i = \gamma^0 \gamma^i$, $\alpha^4 = 0$, $\beta = \gamma^0$.
In this section we look for all possible choices of
$\al^i$, $\al^4$, $\bt$ such that the components of $\Psi$ satisfy
the $\kappa$-deformed KG equation, {\it i.e.} such that a plane
wave which obeys the dispersion relation (\ref{casimir}) \be
\cosh(\lambda E)-\frac{\lambda^2}{2}e^{\lambda
E}\vp^2=\cosh(\lambda m) \ee is a solution.

One can derive the Dirac equation in the familiar commutative
Minkowski spacetime by following the same strategy (indeed Dirac
obtained his equation in this way, by introducing some matrix
coefficients of the elements of the differential calculus and
imposing that these matrices be consistent with the KG equation).
While in commutative Minkowski spacetime the consistency with the
KG equation is sufficient to fully determine the Dirac equation,
in \kM\ this procedure only allows to determine the (deformed)
Dirac equation ``on shell" (one of the matrices that gives the
general parametrization of the deformed Dirac equation remains
undetermined, but it does not affect the form of the equation once
the on-shell dependence of energy on momentum is imposed).
In order to fully determine the equation
it turns out to be necessary to impose a suitable condition
of covariance under the action of the symmetry group (the
covariance under symmetry-group transformations is instead
automatically satisfied in the commutative-Minkowski case, once
the consistency with the KG equation is imposed).

We start by making use of the requirement of consistency with the
commutative ($\lambda\to0$) limit (\ref{diraccomm}): \be \ts
i\D_0+i\al^j\D_j+\al^4\D_4+\bt M_D\td\Psi(\x) =0\rightarrow \ts
i\partial_0+i\gamma^0\gamma^j\partial_j +\gamma^0m\td \psi(x)=0
\label{goodlimit} \ee where $\gamma^{\mu}$ are the familiar Dirac
matrices satisfying the Clifford-algebra relations
$\{\gamma^{\mu},\gamma^{\nu}\} = 2\eta^{\mu\nu}$. From
(\ref{goodlimit}) one easily finds that \bea
\lim_{\lambda=0}\al^i&=&\gamma^0\gamma^i\nn\\
\lim_{\lambda=0}\bt&=&\gamma^0\nn\\
\lim_{\lambda= 0}M_D(m)&=&m\label{classlim} \eea and that the
$\lambda\to0$ limit of $\al^4$ must be finite (if the
$\lambda\to0$ limit of $\al^4$ is not singular the term in the
Dirac equation that comes from the fifth ``spurious" element of
the differential calculus disappears, as needed, in the
$\lambda\to0$ limit).

%\footnote{{\bf spiegazione da togliere nella vrsione finale}\\
%perche' imponendo il limite classico abbiamo che
%\be
%\al^4(0)\lambda^{-1}[-\lambda^2(P_0^2-P^2)+...]+\bt(0)M_D=-\gamma^0 m
%\ee
%ma il primo pezzo deve annullarsi perche'
%dip da $P_0,P$, quindi $al^4=O(1)$ in $\lambda$.}

Let us also observe that, whereas in commutative Minkowski one can
safely assume that the entries of the matrices are just constant numbers
(independent of any of the variables that characterize the system),
the presence
of the scale $\lambda$ in \kM\ forces us to allow for a possible
dependence of the $\al^{\mu},\bt$ on the mass $m$ ($\al^i =
\al^i(\lambda m)$, $\bt = \bt(\lambda m)$, $\al^4 = \al^4(\lambda
m)$). (We therefore allow for such a dependence, but eventually we find
that it is not present.)

Next we observe that, by multiplying Eq.~(\ref{pde:dirac}) by the
operator $i\D_0-(i\al^j\D_j+\al^4\D_4+\bt M_D)$ \be \ts
i\D_0-(i\al^j\D_j+\al^4\D_4+\bt M_D)\td\ts i\D_0
+i\al^j\D_j+\al^4\D_4+\bt M_D\td \Psi(\x)=0 \ee one obtains \be
[\D^2_0+\ts i\al^j\D_j+\al^4\D_4+\bt M_D\td^2]\Psi(\x)=0 ~.
\label{square} \ee

The requirement ii) of consistency with the deformed KG equation
translates into the condition that a choice of $\Psi$ given by
on-shell plane waves (\ref{decomp})\be
u(\vec{p})e^{-ip_j\x_j}e^{iE\x_0}+v(\vec{p})e^{-iS(p_j)\x_j}e^{iS(E)\x_0},
\ee  with $E=E(p)$ given by $\cosh(\lambda
E)-\frac{\lambda^2}{2}e^{\lambda E}\vp^2 =\cosh(\lambda m)$),
should be a solution of our sought deformed Dirac equation. This
allows us to obtain from (\ref{square}) the following $n-plet$ of
equations for $u(\vec{p})$: \bea
 [\D^2_0(E,p)&-&(\al^j)^2(\D_j(E,p))^2+(\al^4)^2\D_4^2+\bt^2M_D^2+ \nonumber\\
&-&\sum_{j<l}\{\al^j,\al^l\}\D_j(E,p)\D_l(E,p)+[i\{\al^j,\al^4\}\D_4
+iM_D\{\al^j,\bt\}]\D_j(E,p)+\nonumber\\
&+&M_D \{\al^4,\bt\}\D_4]u(\vec{p})=0 \label{condizione} \eea

Using the fact that ordinary space-rotation symmetry should still
be preserved the equations (\ref{condizione}) straightforwardly
lead to the consistency requirements:
 \bea
&&\sum_{j<l}\{\al^j,\al^l\}\D_j(E,p)\D_l(E,p)=0\\
&&[\{\al^j,\al^4\}\D_4+M_D\{\al^j,\bt\}]\D_j(E,p)=0\\
&&\sum_{j} (\al^j)^2 (D_j(E,p))^2 \propto \sum_{j} (D_j(E,p))^2
\eea From this we conclude that \bea
\{\al^i,\al^j\}&=&0\;\; i\neq j\label{constr1}\\
\{\al^i,\al^4\}\D_4(k)&=&-M_D\{\al^i,\bt\}\label{constr2}\\
(\al^1)^2&=&(\al^2)^2 = (\al^3)^2\label{constr2bis} \eea We can
now use these results to write equation (\ref{condizione}) as
follows: \be
 [\D^2_0(E,p)-|\vec{\D}(E,p)|^2 {\cal G}+\D_4^2(\al^4)^2
 +M^2_D\bt^2+M_D\{\al^4,\bt\}\D_4]u(\vec{p})=0 \label{eq:KG}
\ee where we introduced the notation ${\cal G}$ for the common
value (see (\ref{constr2bis})) of the $(\al^i)^2$ matrices, ${\cal
G} \equiv (\al^1)^2 =(\al^2)^2 = (\al^3)^2$.

Next we can use the fact that, on the basis of
(\ref{5Ddiffcalc1}), we know that the $D_a(k)$ have the following
on-shell expressions: \bea
\D_0(E,p)&=&\frac{i}{\lambda}[e^{\lambda E}-\cosh(\lambda m)]\nonumber\\
\D_j(E,p)&=&ie^{\lambda E}p_j\nonumber\\
\D_4(E,p)&=&\frac{1}{\lambda}[\cosh(\lambda m)-1] ~. \eea This
allows us to rewrite (\ref{eq:KG}) as \bea
&-&\lambda^{-2}[e^{2\lambda E}+\cosh^2(\lambda m)
-2\cosh(\lambda m)e^{\lambda E}]I+\nonumber\\
&+&{\cal G} e^{2\lambda E}|\vp|^2+\nonumber\\
&+&(\al^4)^2\lambda^{-2}(1-2\cosh(\lambda m)+\cosh^2(\lambda m))+M^2_D\bt^2+\\
&+&\lambda^{-1}M_D\{\al^4,\bt\}(\cosh(\lambda m)-1)=0 \nonumber ~,
\eea which can also be cast in the form
%\footnote{metto in nota i passaggi,
%penso si possano omettere. Si riscrive la relaz di disp:
%\bea
%2(\cosh(\lambda E)-\cosh(\lambda m))
%=\lambda^2p^2e^{\lambda E}&&2(e^{\lambda E}\cosh(\lambda E)-
%e^{\lambda E}\cosh(\lambda m))=\lambda^2p^2e^{2\lambda E}\nn\\
%(e^{2\lambda E}+1-2e^{\lambda E}\cosh(\lambda m))=\lambda^2p^2
%e^{2\lambda E}&&(e^{2\lambda E}+\cosh(\lambda m)^2-2
%e^{\lambda E}\cosh(\lambda m))=\lambda^2p^2e^{2\lambda E}-\sinh(\lambda E)^2\nn
%\eea
%quindi si sostituisce questo nell'equazione}
\bea &&-\sin^2(\lambda m)I+\lambda^2[{\cal G}-I]e^{2\lambda
E}|\vp|^2
+\nonumber\\
&+&(\al^4)^2(1-2\cosh(\lambda m)+\cosh^2(\lambda
m))+\lambda^2M^2_D\bt^2
+\nonumber\\
&+&\lambda M_D\{\al^4,\bt\}(\cosh(\lambda m)-1)=0 ~, \label{joc61}
\eea using again the dispersion relation.

Since Eq.(\ref{joc61}) must hold for every arbitrary value of
$\vp$ we can deduce that \be {\cal G} = I \ee and \be
\sinh^2(\lambda m)I-(\al^4)^2(1-\cosh(\lambda m))^2
-\lambda^2M_D^2\bt^2+M_D\lambda\{\al^4,\bt\}(1-\cosh(\lambda m))=0
\ee which can be conveniently rewritten as \be \sinh^2(\lambda
m)I=(\al^4(\cosh(\lambda m)-1)+\lambda M_D \bt)^2 ~.
\label{constr3} \ee

At this point we have reduced our search of consistent deformed
Dirac equations to the search of matrices $\{ \al^j,\al^4, \beta
\}$ such that the following requirements
(\ref{constr1}-\ref{constr2}-\ref{constr3}) are satisfied: \bea
\{\al^j,\al^k\} &=& 2\delta^{jk} I\nn\\
\{\al^j,\al^4\}[\cosh(\lambda m)-1]&=&-\lambda M_D \{\al^j,\bt\}
\label{const2}\\
\sinh^2(\lambda m)I&=&(\al^4(\cosh(\lambda m)-1)+\lambda M_D
\bt)^2 \label{const3} \eea

In deriving from these requirements an explicit result for the
matrices $\{ \al^j,\al^4, \beta \}$ it is convenient to first
consider the case in which $M_D m\neq0$, the case of massive
particles. It is then convenient
to introduce the matrix $A$ \be A \equiv
{\left(\al^4(\cosh(\lambda m)-1)+\lambda M_D \bt \right) \over
\sinh(\lambda m)} \label{A} \ee which allows us to cast the deformed
Dirac equation in the following form:
 \be
[i\D_0(k)+i\al^j\D_j(k) + \D_4(k)\frac{\sinh(\lambda
m)}{\cosh(\lambda m)-1}A+M_D[-\frac{\lambda\D_4(k)}{\cosh(\lambda
m)-1} +1]\bt]\tilde{\psi}(k)=0 \label{jodirac1} \ee From
(\ref{const2}) and (\ref{const3}) it follows that $\{A,\al^i\}=0$
and $A^2=I$.

We are at this point ready to obtain the most general DSR-deformed
Dirac equation in \kM. In fact, Eq.~(\ref{jodirac1}) for the
(on-shell) spinor $u(\vec{p})$ (\ref{decomp}) simplifies to \be
[i\D_0(E,p)+i\D_j(E,p)\al^j +\frac{\sinh(\lambda
m)}{\lambda}A]u(\vec{p})\label{7} \ee where the matrices $\al^j,A$
satisfy the conditions\footnote{From these conditions one can also
infer that the $n {\times} n$ matrices we are seeking must have
$n$ even and $n \ge 4$ (not smaller than $4 {\times} 4$ matrices).
In fact, from the anticommutation relations it follows that
$TrA=0$, $A^2=1$, and $detA={\pm}1$ which requires $n$ to be even.
The case $n=2$ is also excluded since there are only $3$
independent anticommuting $2 {\times} 2$ matrices (Pauli
matrices). We take $n=4$ just as in the $\lambda \rightarrow 0$
(commutative-Minkowski) limit.} \be \{\al^j,\al^k\}=2\delta^{jk}
I,\;\;\;,\{\al^j,A\}=0,\;\;\;A^2=I \label{alfajo} \ee Introducing
$\gamma^0 \equiv A$ and $\gamma^j \equiv A\al^j$ one finds from
(\ref{alfajo}) that the $\gamma$'s must satisfy \be
\{\gamma^{\mu},\gamma^{\nu}\}=2\eta^{\mu \nu},\;\;\;\mu=0,1,2,3.
\ee {\it i.e.} they must be the usual (undeformed!) Dirac
matrices. And we find (by multiplying (\ref{7}) by $A$ and making
use of  $\gamma^0 \equiv A$ and $\gamma^j \equiv A\al^j$) that in
terms of the usual Dirac matrices there is a unique solution to
our problem of finding the most general DSR-deformed Dirac
equation in \kM : \be [\frac{e^{\lambda E}-\cosh(\lambda
m)}{\lambda}\gamma^0+p_j e^{\lambda E}\gamma^j-\frac{\sinh(\lambda
m)}{\lambda} I]u(\vec{p})=0\label{on-shell} \ee It is rather
satisfactory that there is a unique deformed Dirac equation in \kM
, and that it reproduces the deformed Dirac equation which had
been already derived, without any {\it a priori} assumption about
the spacetime structure, in Section~2 .

\subsection{Invariance under \kkP\ action}
We established that the Dirac equation in \kM\ must take
the form (\ref{on-shell}) ``on shell": our analysis so far
has allowed us to fully determine the form of the Dirac equation
with the exception of the matrix $\bt$ which is still undetermined,
but $\bt$ does not affect the form of the equation once
the on-shell dependence of energy on momentum is imposed.
In order to
determine the matrix $\bt$ it is necessary to impose the
covariance of our deformed Dirac equation in the \kkP\ sense.
This will require us to introduce a representation of the
$\kappa$-Poincar\'{e} (Hopf) algebra for spin-$1/2$ particles.

In preparation for this analysis we first briefly review for the
reader the analogous analysis for the classical Poincar\'{e} (Lie)
algebra. A key point is that for the Lorentz sector the
representation can be described as the sum of two parts: \be
M^T_j=M_j+m_j\;\;\;N^T_j=N_j+n_j \ee
%\be
%J_{\mu\nu}^T=X_{\mu\nu}+\Sigma_{\mu\nu}
%\ee
where $(M_j,N_j)$ is a spinless unitary representation of $O(3,1)$
and acts in the "outer" space of the particle (the one the
codifies the momenta and orbital momenta of the particle), whereas
$(m_j,n_j)$ is a  finite-dimensional representation of $O(3,1)$
and acts in the "inner" space (spin indices).
%\be
%[J_{\mu\nu}^T,J_{\rho,\sigma}^T]=i[\eta_{\nu\rho}J_{\mu\sigma}^T-\eta_{\nu\sigma}J_{\mu\rho}
%^T-\eta_{\mu\nu}J_{\nu\sigma}^T+\eta_{\mu\sigma}J_{\nu\rho}^T]
%\ee
A representation of the whole Poincar\'{e} group is then obtained
by introducing four translation generators
$P_{\mu}=-i\partial_{\mu}$ that act only in the "outer" space of
the particle. The spinorial representation of the classical
Poincar\'{e} algebra is therefore given by \be
P_{\mu}^T=P_{\mu}\;\;\;M^T_j=M_j+m_j\;\;\;N^T_j=N_j+n_j \ee which
of course satisfy the following familiar commutation relations:
\bea
&&[P_\mu^T,P_\nu^T]=0 \nn\\
&& \qs M_j^T,M_k^T\qd=\!\! i\epsilon_{jkl}M_l^T ~,~~~[N_j^T,N_k^T]
=-i\epsilon_{jkl}M_l^T ~,~~~[M_j^T,N_k^T]=i\epsilon_{jkl}N_l^T\nn\\
&&\qs M_j^T,P_0^T\qd=0\;\;\;\;\; [M_j^T,P_k^T]=i\epsilon_{jkl}P_l^T\nn \\
&&\qs N_j^T,P_0^T\qd=iP_j^T
\;\;\;\;\;[N_j^T,P_k^T]=i\delta_{jk}P_0^T\nn \eea
%\bea
%[J_{\mu\nu}^T,J_{\rho,\sigma}^T]&=&i[\eta_{\nu\rho}J_{\mu\sigma}^T-\eta_{\nu\sigma}J_{\mu\rho}
%^T-\eta_{\mu\nu}J_{\nu\sigma}^T+\eta_{\mu\sigma}J_{\nu\rho}^T]\nn\\
%\qs J_{\nu\rho}^T,P_{\mu}\qd&=&i(\eta_{\mu\rho}P_{\nu}-\eta_{\mu\nu}P_{\rho})\nn
%\eea
The differential form of the spinless realization is given by: \be
P_\mu=-i\partial_\mu\;\;M_j=\epsilon_{jkl}x_kP_l\;\;N_j=x_jP_0-x_0P_j
\ee and the finite dimensional realization can be expressed in
terms of the familiar $\gamma$ matrices \be
m^j=\frac{i}{4}\epsilon^{jkl}\gamma^k\gamma^l ~,~~~ n^j
=\frac{i}{2}\gamma^j\gamma^0 \ee The action of the global
generators of Poincar\'{e}  over a Dirac spinor is: \bea
M^T_j&&\psi_r(x)=\int d^4k\; [(M_j e^{ik x} )\tilde{\psi}_r(k)
+ e^{ik x} (m_j\tilde{\psi}_r(k))]\nn\\
N^T_j&&\psi_r( x)=\int d^4k\; [(N_j e^{ik x} )\tilde{\psi}_r(k)
+ e^{ik x} (n_j\tilde{\psi}_r(k))]\nn\\
P_{\mu}^T&&\psi_r( x)=\int d^4k\; [(P_{\mu} e^{ik x})
\tilde{\psi}_r(k)]\nn \eea
%At this point one can introduce a Hopf algebra language in order to express the action of %Poincar\'{e} over a tensor product of the two space of the particle (spatial and spin). In this
% optics we introduce the coproduct that is useful in order to compose representations (the
% action is essentially a representation)\footnote{to be rigorous these coproducts come from
%the global representation of de Sitter algebra $S0(3,2)$ that through a contraction procedure
%is reduced to $IO(3,1)$ Poincar\'{e} algebra}:
%\bea
%\Delta (M_j)&=&M_j\otimes 1+1\otimes M_j\nn\\
%\Delta (N_j)&=&N_j\otimes 1+1\otimes N_j\nn\\
%\Delta (P_{\mu})&=&P_{\mu}\otimes 1+1\otimes P_{\mu}\nn
%\eea
%In the case of a simpe Lie Poincar\'{e} algebra the coproducts are trivial.
%We can express the action of the global generators by the coproduct:
%\bea
%(M,N)^T&{\cdot}&\psi_r(\x)=\int dk\;\Delta (M,N){\cdot} [:e^{ik\x}:\otimes \tilde{\psi}_r(k)]\nn\\
%P_i^T&{\cdot}&\psi_r(\x)=\int dk\;\Delta (P_i){\cdot}[:e^{ik\x}:\otimes \tilde{\psi}_r(k)]\nn
%\eea
%The invariance condition of th eDirac equation translate in the conditions:
%\be
%[J_{\mu\nu}^T,\Dirac_{\lambda}]=0
%\ee
%The solution for the classical global generators results to be:
%utili le seguenti relazioni di commutazione:
%\bea
%[n^i,\gamma^0]&=&i\gamma^i\nn\\
%\qs n^i,\gamma^k\qd&=&i\gamma^0\delta^{ik}\nn\nn\\
%\qs m^i,\gamma^0\qd&=&0\nn\\
%\epsilon^{ijk}\qs m^k,\gamma^l\qd P^jP^l&=&i[\gamma^iP^2-\gamma^jP^jP^i]
%\eea
and the Dirac operator is of course an invariant: \be
[P_{\mu}^T,\Dirac]=[M_j^T,\Dirac]=[N_j^T,\Dirac]=0 \ee

We intend to obtain analogous results for spinors and the Dirac
operator in \kM. Our deformed Dirac operator must be invariant,
\be [{\mathcal{P}}_{\mu}^T,\Dirac_{\lambda}]
=[{\mathcal{M}}_j^T,\Dirac_{\lambda}]
=[{\mathcal{N}}_j^T,\Dirac_{\lambda}]=0 \ee under the action of
\kkP\ generators
${\mathcal{P}}^T,{\mathcal{M}}^T,{\mathcal{N}}^T$, which satisfy
the following commutation relations: \bea
\qs{\mathcal{P}}_{\mu}^T,{\mathcal{P}}_{\nu}^T\qd&=&0\nn\\
\qs{\mN}_j^T,{\mN}_k^T\qd&=&-
i\epsilon_{jkl}{\mM}_l^T\;\;\;\qs{\mN}_j^T,{\mM}_k^T \qd=
i\epsilon_{jkl}{\mN}_l^T\nn\\
\qs{\mM}_j^T,{\mM}_k^T \qd&=&
i\epsilon_{jkl}{\mM}_l^T\nn\\
\qs{\mM}_j^T,{\mathcal{P}}_0^T\qd&=&0\;\;\;\qs{\mM}_j^T,{\mathcal{P}}_k^T\qd
=i\epsilon_{jkl}{\mathcal{P}}_l^T\nn\\
\qs{\mN}_j^T,{\mathcal{P}}_0^T \qd&=&
iP_j^T\;\;\;\qs{\mN}_j^T,{\mathcal{P}}_k^T\qd=i\delta_{jk}[\frac{1-e^{-
2\lambda{\mathcal{P}}_0^T}}{2\lambda}+\frac{\lambda}{2}({{\mathcal{P}}^T})^2]
-i\lambda{\mathcal{P}}_j^T{\mathcal{P}}_k^T\nn \eea

Consistently with the results obtained so far we expect that it
will not be necessary to deform the rotations: \be {\mM}^T_j =
M^T_j = M_j+m_j \ee and in fact this satisfies all consistency
requirements, as one can easily verify.

Boosts in general require a deformation, and we already know from
the earlier points of our analysis that the differential form of
the spinless realization must be given by the operators ${\mN}_j$
(see (\ref{bicrossbasis})). We therefore need a suitable
finite-dimensional realization ${\tilde n}_j$ of boosts, so that
${\mN}_j^T$ will be given by ${\mN}_j^T = {\mN}_j + {\tilde n}_j$.
Using the fact that \be [{\mN}_j,\D_0]=i\D_j,\;\;\;\qs
{\mN}_j,\D_k\qd=i\D_0\delta_{jk},\;\;\;\qs {\mN}_j,\D^4\qd=0 \ee
it is easy to verify\footnote{Details of this lengthy, but rather
straightforward, analysis will be soon available on the
arXiv~\cite{alesstesi}.} that with ${\tilde n}_j = n_j$ and
$\bt=0,\gamma^0,\gamma^0\gamma^5$ one has a form of  ${\mN}_j^T$
\bea {\mathcal{N}}_j^T={\mathcal{N}}_j+n_j \label{gac789} \eea
which satisfies (with ${\mathcal{P}}^T$ and ${\mathcal{M}}^T$) the
Hopf-algebra requirement, and a Dirac equation which is invariant
under these Hopf-algebra transformations. We will therefore
describe ${\mN}_j^T$ with (\ref{gac789}) and we could consider
three possibilities for the matrix $\bt$ ($\bt=0$, $\bt=\gamma^0$
and $\bt=\gamma^0\gamma^5$).

Actually only $\bt=\gamma^0$ is acceptable; in fact, both for
$\bt=0$ and for $\bt=\gamma^0\gamma^5$ it is easy to check that
our deformed (off-shell) Dirac equation would not reproduce the
correct $\lambda \rightarrow 0$ (classical-spacetime) limit.

%In the case of $\bt=\gamma^0$, making the limit
%and remembering (\ref{classlim}):
%\bea
%&&\qs \gamma^0P_0+\gamma^iP_i-m[\frac{\lambda \D_4(P)}{1
%-\cosh(\lambda m)}+1]+\D_4(P)\frac{\sinh(\lambda m)}{1
%-\cosh(\lambda m)}I\qd \psi({\x})=0\nn\\
%&&\qs \gamma^0P_0+\gamma^iP_i-m[\frac{\lambda \D_4(P)}{1
%-\cosh(\lambda m)}+1]+\lambda\D_4(P)\frac{m}{1-\cosh(\lambda
%m)}I\qd \psi({\x})=0\nn\\
%&&\qs \gamma^0P_0+\gamma^iP_i-mI\qd \psi({\x})=0\nn\\
%\eea
%but in the case of $\bt=0$ the classical limit is not satisfied, in fact:
%\bea
%&&\qs \gamma^0P_0+\gamma^iP_i+\D_4(P)\frac{\sinh(\lambda m)}{1
%-\cosh(\lambda m)}I\qd \psi({\x})=0\nn\\
%&&\qs \gamma^0P_0+\gamma^iP_i+m\frac{\lambda^2(P_0^2-P^2)
%+O(\lambda^4)}{-\lambda^2m^2+O(\lambda^4)}I\qd \psi({\x})=0\nn\\
%\eea
%this equation is the classical Dirac equation only if $P_0^2=P^2+m^2$
%(che e' la condizione on shell nel limite classico ma noi stiamo
%considerando il caso fuori dalla shell, quindi a me sembra che
%dobbiamo scartare questa soluzione).

%It is very easy to cheque that olso
%with $\bt=\gamma^0\gamma^5$ we cannot obtain the correct commutative limit.

We are left with a one-parameter family ($M_D$ is the parameter)
of deformed Dirac equations \be \qs
i\gamma^0\D_0(P)+i\gamma^j\D_j(P) +(\D_4(P)\frac{\sinh(\lambda
m)-\lambda M_D}{\cosh(\lambda m)-1}+M_D)I\qd \Psi({\x})=0
\label{final5D} \ee As far as we can see the free parameter $M_D$
does not have physical consequences (it clearly does not affect
the on-shell equation), and it appears legitimate to view it as a
peculiarity associated with the nature of the (five-dimensional)
differential calculus in \kM. The choice $M_D =m$ is
allowed (but not imposed upon us) by the formalism, which
actually allows $M_D =m f(\lambda m)$ with
any $f$ such that  $M_D \rightarrow m$ for $\lambda \rightarrow 0$
and $M_D \rightarrow 0$ for $m \rightarrow 0$.
In addition to $M_D =m$, other noteworthy possibilities
are $M_D=\frac{\sinh(\lambda m)}{\lambda}$, which
corresponds to the \emph{ansatz} $\al^4=0$ of Section~3.3,
and $M_D = M_{KG}=\sqrt{2(\cosh(\lambda m)-1)}/\lambda$.
Since $M_D$ does not affect the on-shell equation, our conclusion
that the on-shell Dirac equation in \kM\ reproduces the on-shell
Dirac equation obtained, using only DSR criteria, in Section~2 is
independent of this freedom for the parameter $M_D$.

\subsection{Massless particles}
Since the on-shell Dirac equation in \kM\ is just the one already
obtained in Section~2, clearly the case of on-shell massless
particles ($m \rightarrow 0$) in \kM\ is also consistent with the
corresponding result already discussed in Section~2.

Concerning a space-time formulation of the deformed Dirac equation
for massless particles we simply observe that (\ref{final5D}) has
a well-defined $m \rightarrow 0$ limit: \be \qs
i\gamma^0\D_0(P)+i\gamma^j\D_j(P) \qd \Psi({\x})=0
\label{final5Dnomass} \ee which is therefore well suited for the
description of massless spin-$1/2$ particles.

\subsection{Aside on a possible ambiguity in the derivation of the
Dirac equation in \kM} When we introduced
 \bea
dF({\x})=dx^a{\D}_a(P)F({\x})\;\;\;\;a=0,\dots,4 \eea with \bea
{\D}_a(P)=\ts\frac{i}{\lambda}[\sinh(\lambda P^0)
+\frac{\lambda^2}{2}e^{\lambda P^0}P^2],iP_j e^{\lambda
P_0},-\frac{1}{\lambda}(1-\cosh(\lambda P_0)
+\frac{\lambda^2}{2}P^2 e^{\lambda P_0})\td\nn\\
\eea we overlooked an equally valid way of introducing the
exterior derivative operator $d$ of a generic \kM\ element
$F({\x})=\Omega(f(x))$ in terms of the 5D differential calculus:
 \bea
dF({\x})=\bar{\D}_a F({\x}) dx^a \;\;\;\;a=0,\dots,4 \eea with
\bea \bar{\D}_a(P)=\ts \frac{i}{\lambda}[\sinh(\lambda P^0)
-\frac{\lambda^2}{2}e^{\lambda P^0}P^2],iP_j,
\frac{1}{\lambda}(1-\cosh(\lambda P_0)
+\frac{\lambda^2}{2}P^2 e^{\lambda P_0})\td\nn\\
\eea

There is however a simple relation between $\D(k)$ and
$\bar{D}(k)$ deformed derivatives: \be \bar{\D}(k)=-\D(S(k))
\;\;\;S(k)=(-k_0,-e^{\lambda k_0}k_j)\label{relaz} \ee (where $S$
is the antipode map, which generalizes the inversion operation in
the way that is appropriate for \kM\
studies~\cite{kpoinap,gacmaj}), and the careful reader can easily
verify\footnote{Details of this straightforward, but tedious,
analysis will be soon available on the arXiv~\cite{alesstesi}.}
that there is no real ambiguity due to the choice of formulation
of the exterior derivative operator $d$. The same physical Dirac
theory is obtained in both cases.

\section{An obstruction for a Dirac equation in \kM\ based
on a four-dimensional differential calculus} In alternative to the
five-dimensional calculus which we have so far considered some
studies (see, {\it e.g.}, Ref.~\cite{gacmaj}) of  \kM\ spacetime
have used a four-dimensional differential calculus\footnote{This
four-dimensional differential calculus was originally
obtained~\cite{majoeck} as a generalization of a two-dimensional
differential calculus over two-dimensional \kM .}: \be
[{\x}_{\mu},dx_j]=0, \;\;\;[x_{\mu},dx_0]=i\lambda dx_{\mu} \ee
One can then express the derivative operator of the element
$\Psi({\x})$ of \kM\ in the following way: \be
d\Psi=\tilde{\partial}_{\mu}\Psi({\x}) dx^{\mu} \ee where
$\tilde{\partial}_{\mu}$ are deformed derivatives that act on the
time-to-the-right-ordered exponential as follows: \bea
&&\tilde{\partial}_je^{ik_0{\x}_0}e^{-ik{\x}}
=\partial_je^{ik_0{\x}_0}e^{-ik{\x}}=
i k_je^{ik_0{\x}_0}e^{-ik{\x}}\nn\equiv d_j(k)e^{ik_0{\x}_0}e^{-ik{\x}}\\
&&\tilde{\partial}_0e^{ik_0{\x}_0}e^{-ik{\x}}=\frac{i}{\lambda}(1-
e^{-\lambda k_0})e^{ik_0{\x}_0}e^{-ik{\x}}\equiv
d_0(k)e^{ik_0{\x}_0} e^{-ik{\x}}\nn \eea

It has been previously established~\cite{gacmaj} that it
is possible to formulate the deformed Klein-Gordon
equation (\ref{defKGeq})
in terms of this four-dimensional calculus:
\be
[\tilde{\partial}^\mu\tilde{\partial}_\mu L+M^2_{KG}]\Phi(\x)
\label{jocKG} \ee
where $L$ is the shift
operator $L\Phi(\vec{\x},\x_0)=\Phi(\vec{\x},\x_0-i\lambda)=
e^{i\lambda \partial_0}\Phi(\x)$. In fact, one easily
finds that $\tilde{\partial}^\mu\tilde{\partial}_\mu
L=\D^a\D_a=\Box_\lambda $.

On the basis of the fact that one can write the deformed Klein-Gordon
equation (\ref{defKGeq}) equivalently
in terms of the four-dimensional
calculus and the five-dimensional calculus, one could guess that
these two examples of differential calculus are
equally well suited for implementing the DSR principles in \kM .
However, we find that this is not
the case. The richer structure of the Dirac equation is more
sensitive to the details of the differential calculus, and the
choice of the five-dimensional differential calculus turns out to
be most natural.

In support of this observation let us attempt
to proceed with the four-dimensional calculus just
as done for the five-dimensional calculus: we write a general
parametrization of a deformed Dirac equation, \be \ts
id_0(k)+id_j(k) \rho^j+M_D' \sigma \td \tilde{\psi_{\kappa}}(k) =0
\label{dirack4} \ee where $\rho^i,\sigma$ are four matrices
(constant or at most dependent on $\lambda m$) to be determined by
imposing that an on-shell ``plane wave" (\ref{decomp}) (with
$\cosh(\lambda E)-\lambda^2e^{-\lambda E}k^2=\cosh(\lambda m)$) is
solution of the deformed Dirac equation and by imposing covariance
in the \kkP\ sense.

The requirement that an on-shell plane wave is a solution leads to
\bea
 &&\qs d^2_0(E,p)-(\rho^j)^2(d_j(E,p))^2+{M_D'}^2\sigma^2+\right.\nonumber\\
&&-\left.\sum_{j<k}\{\rho^j,\rho^k\}d_j(E,p)d_k(E,p)+M_D'\{\rho^j,\sigma\}
d_j(E,p)\qd u(\vec{p})=0 \label{condizione4} \eea from which one
derives as necessary conditions: \be
\{\rho^j,\rho^k\}=0,\;\;\{\rho^j,\sigma\}=0,\;\;
(\rho^1)^2=(\rho^2)^2=(\rho^3)^2 \label{joc64} \ee These
conditions are necessary but not sufficient, and actually there is
no choice of the matrices $\rho^j,\sigma$ of the type that we are
seeking that allows to satisfy (\ref{condizione4}) for all values
of the momentum $p$. To see this let us use (\ref{joc64}) to
rewrite (\ref{condizione4}) as \be (d_0^2(E,p)I-d_j(E,p)^2 {\cal
Q}+{M_D'}^2\sigma^2)u(\vec{p})=0 \label{con} \ee where ${\cal Q}
\equiv (\rho^1)^2=(\rho^2)^2=(\rho^3)^2$. In this Eq.~ (\ref{con})
we are left with two unknown matrices, ${\cal Q},\sigma$, to be
determined, and it is easy to see that there is no choice of
${\cal Q},\sigma$ that allows to satisfy (\ref{con}) for all
values of the momentum $p$. For example, by looking at the form of
the equation for $p=0$ (and $E=m$) one is forced to conclude that
\be \sigma^2= - d_0^2(m)I/{M_D'}^2(m) \ee but then, with this
choice of $\sigma^2$, Eq.~(\ref{con}) turns into an equation for
${\cal Q}$ which does not admit any solution of the type we are
seeking: \be (d_0^2(E)-d_0^2(m)-d_j^2 {\cal Q})=-(1-e^{-\lambda
E})^2 +(1-e^{-\lambda m})^2+\lambda^2p^2 {\cal Q}=0 \ee {\it i.e.}
(using again the dispersion relation) \be {\cal Q} =
[(1-e^{-\lambda E})^2-(1-e^{-\lambda m})^2][( e^{2\lambda
E}+1)-2e^{\lambda E}\cosh(\lambda m)]^{-1} \ee

What we have found is that there is no choice of
energy-momentum-independent matrices $\rho^j,\sigma$ that can be
used in order to obtain a consistent Dirac equation for \kM . The
analogous problem for the 5D calculus did have a perfectly
acceptable solution. Here, with the four-dimensional differential
calculus, we would be led to consider energy-dependent matrices
$\rho^i,\sigma$ but this is unappealing on physical grounds and in
any case the fact that this awkward assumption can be avoided in
the five-dimensional calculus appears to be a good basis for
preferring the five-dimensional calculus over the four-dimensional
calculus.

\section{Comparison with previous results on deformed Dirac equations}
Our main objective was to formulate a DSR-deformed Dirac equation,
and we showed that given the formulation of a DSR proposal in
energy-momentum space one can very straightforwardly obtain a
corresponding modification of the Dirac equation in
energy-momentum space. We then used our result on DSR-deformed
Dirac equation as an opportunity to explore the hypothesis that
\kM\ spacetime, at least the one-particle sector of theories in
\kM\ spacetime (see Subsection~3.1), might provide a spacetime
arena for the DSR framework. From this perspective it was a key
point for us to investigate under which assumptions about the
formulation of spin-$1/2$ particles in \kM\ one would reobtain the
DSR-deformed Dirac equation. We considered alternative
formulations of the differential calculus, and for each given
choice of differential calculus we were interested in finding the
most general compatible formulation of the Dirac equation. This
allowed us to establish which choice of differential calculus was
required for a DSR-compatible result, and to investigate whether
in addition to the choice of differential calculus there were
other choices to be made in order to obtain the DSR-deformed Dirac
equation. We found that the ``5D" differential calculus should be
preferred to its ``4D" alternative, and we found that, once the
choice of the 5D differential calculus is made, one is then led
automatically to the DSR-deformed Dirac equation.

While there was no previous study of deformed Dirac equation from
the DSR perspective, there has been some previous work of a
deformed Dirac equation for \kM\ and on a deformed Dirac equation
governed by the structure of the \kkP\ Hopf algebra. We find
appropriate to comment here on these related studies.

One early investigation of a deformed Dirac equation governed by
the structure of \kkP\ was reported in Ref.~\cite{LNR92}, where
however the momentum generators were described as differential
generators on a commutative spacetime.

The study reported in Ref.~\cite{Firenzegroup} adopts a general
perspective which is closer to ours, but it focused on the
so-called standard basis of \kkP\ for which (unlike the case we
considered of the Majid-Ruegg bicrossproduct basis) it remains
unclear~\cite{dsrnext,dsrIJrev} whether a connection with the
DSR criteria can be established. For the standard basis there is
not even a clear connection with a choice of ordering convention
for functions in \kM , whereas in our analysis the connection
between the Majid-Ruegg bicrossproduct basis and the
time-to-the-right ordering convention played a key role. Moreover
it was important for our line of analysis to consider that the
construction of the deformed Dirac equation in full generality
(including, for example, the search of acceptable forms for the
matrix $\alpha^4$ and for the $M_D$ parameter) whereas in
Ref.~\cite{Firenzegroup} a more limited class of possibilities was
considered.

Ref.~\cite{Bibikov} took as starting point the \kM\ spacetime and
considered the construction of a Dirac equation for massless
spin-$1/2$ particles. The perspective is considerably different
from ours, since takes inspiration from the Connes criteria for a
connection between differential operators and the Dirac operator.
Besides the difference in perspective, the fact that
Ref.~\cite{Bibikov} considers only massless spin-$1/2$ particles
reduces its relevance to the problem we considered, where the
particle mass (and its relation to various mass parameters) played
a key role. The careful reader can easily verify that in the
derivation of the Dirac equation following our strategy some of
the conditions are multiplied by the mass parameter, and therefore
those conditions are formally irrelevant in the massless case,
leading to a less constrained framework. We have chosen to
introduce massless particles at the end of the analysis, imposing
continuity of the $m\rightarrow 0$ limit (so that, by continuity,
the relevant conditions encountered for nonvanishing mass are
taken into account also in the massless limit).

In Ref.~\cite{masl2} the emphasis is placed on the structure of
the \kkP\ Hopf algebra and the analysis does not properly consider
\kM\ spacetime. In fact, Ref.~\cite{masl2} introduces a
five-dimensional metric and a fifth spacetime coordinate (which
commutes with the other four \kM -type coordinates). A
corresponding formulation of the Dirac equation is found by
requiring that, in an appropriate sense, the Dirac operator should
be a square root of the Klein-Gordon operator.

In Ref.~\cite{masl3} the analysis does concern \kM\ spacetime. but
the proposed deformation of the Dirac equation is obtained by
enforcing certain criteria based on the search of the unitary
representations of the so-called ``\kkP\ group" with noncommuting
group parameters. The physical meaning of noncommuting group
parameters remains rather obscure\footnote{In conventional
theories, with conventional Lie-group symmetries, on obtains a
group elements by exponentiation of the generators of the algebra,
$e^{a_j Tj}$, with commuting parameters $a_j$. For this situation
the physical interpretation is well digested. It remains to be
established whether a consistent physical interpretation can be
given for the case in which the parameters $a_j$ satisfy
nontrivial algebraic relations (noncommutativity), as in the case
of the ``\kkP\ group" construction.}. And it appears difficult to
establish whether the criteria proposed in Ref.~\cite{masl3} are
as general as ours (we tentatively see one less free matrix
introduced in the initial parametrization of the analysis).

\section{Summary and outlook}
The recent, rather strong, interest in the DSR framework has
focused in part on some experimental contexts in which the
kinematic properties of fundamental particles are analyzed. Some
of these analyses involve spin-$1/2$ particles, but there was no
direct derivation of a DSR-deformed Dirac equation. We have filled
this gap in Section~2, where indeed we showed that a DSR-deformed
Dirac equation can be derived straightforwardly on the basis of
the laws of energy-momentum transformation in the DSR framework.

There has also been interest in the possibility that the \kM\
noncommutative spacetime might provide an example of quantum
spacetime in which DSR symmetries are present. Some difficulties
have been encountered in enforcing DSR symmetries in the
two-particle (and multi-particle) sector of theories in \kM , but
at least in the one-particle sector there is growing evidence of
the connection between DSR and \kM . We provided in Sections~3, 4
and 5 additional evidence of this connection. Our analysis showed
however that it might be improper to state in full generality that
\kM\ spacetime is DSR invariant; in fact, in order to satisfy the
DSR requirements some structures must be introduced consistently
in \kM . A noteworthy example of this fact is provided by the
choice of differential calculus which we stressed. Previous
analyses, focusing on the (deformed) Klein-Gordon equation
appeared to suggest that both the 4D differential calculus and the
5D differential calculus should be equally well suited for the
formulation of DSR-compatible theories in \kM . Our analysis of
the (deformed) Dirac equation, with its richer structure, shows
that this is not the case: only the 5D differential calculus leads
to a Dirac equation which is acceptable from a DSR perspective.

While most of our analysis focused on the DSR perspective and on
the possible role of \kM\ in DSR theories, as we stressed in
Section~6 some of the results we obtained appear to contribute
to the literature on various formulations of the
Dirac equation motivated by \kkP\ and/or \kM .

An interesting issue which could be considered
in future studies is the one of the fate
of Poincar\'{e} symmetries in a \kM\ spacetime equipped
with the ``4D" differential calculus.
This choice of differential calculus does not appear to lead to any
pathologies from the perspective of \kM\ mathematics, but it
is clearly incompatible with classical Poincar\'{e} symmetries
and it also fails to produce a DSR-deformed Dirac equation.

\section*{Acknowledgments}
We are grateful for conversations with F.~D'Andrea during the
transition from the first version of this manuscript
(http://arXiv.org/abs/gr-qc/0207003v1) to the present version (a
transition that lasted a considerable amount of time because of
the desire to include some of the results then in preparation for
Ref.~\cite{alesstesi}.)

\baselineskip 12pt plus .5pt minus .5pt

\end{document}